\documentclass[conference]{IEEEtran}

\IEEEoverridecommandlockouts
\usepackage[utf8]{inputenc}
\usepackage{blindtext}
\usepackage{subcaption}

\usepackage{graphicx}
\graphicspath{{figures/draft/}}

\usepackage{hyperref}
\usepackage{upgreek}
\usepackage{siunitx}
\usepackage{float}
\usepackage{stfloats}
\usepackage{placeins}
\usepackage{balance}
\usepackage{amsmath,amsfonts,amssymb}

\usepackage{ragged2e}

\usepackage{glossaries}

\newcommand*\myglsentry[1]{%
   \glsentrylong{#1}%
}

\newacronym[longplural={Frames per Second}]{fpsLabel}{FPS}{Frame per Second}
\newacronym[longplural={Tables of Contents}]{tocLabel}{TOC}{Table of Content}
\newacronym{act}{ACT}{Asynchronous Circuit Toolkit}
\newacronym{adc}{ADC}{Analog Digital Converter}
\newacronym{aer}{AER}{Address Event Representation}
\newacronym{afe}{AFE}{Analog FrontEnd}
\newacronym{ai}{AI}{Artificial Intelligence}
\newacronym{ampa}{AMPA}{$\alpha$-amino-3-hydroxy-5-methyl-4-isoxazolepropionic acid}
\newacronym{ams}{AMS}{Analog Mixed-Signal}
\newacronym{api}{API}{Application Programming Interface}
\newacronym{bap}{BAP}{back-propagating action potential}
\newacronym{bcall}{BCaLL}{Bi-Stable Calcium based Local Learning}
\newacronym{bd}{BD}{Bundled Data}
\newacronym{beol}{BEOL}{Back-End Of Line}
\newacronym{bjt}{BJT}{Bipolar Junction Transistor}
\newacronym{bnc}{BNC}{Bayonet-Neill-Concelman}
\newacronym{bptt}{BPTT}{Back Propagation Through Time}
\newacronym{bp}{BP}{Back Propagation}
\newacronym{camp}{cAMP}{cyclic Adenosine Mono Phosphate}
\newacronym{cam}{CAM}{Content Addressable Memory}
\newacronym{cco}{CCO}{Current Controlled Oscillator}
\newacronym{cc-by}{CC-BY}{Creative Commons Attribution}
\newacronym{chp}{CHP}{Communicating Hardware Processes}
\newacronym{cl}{CL}{Closed-Loop}
\newacronym{cm}{CM}{Continuum Mechanics}
\newacronym{cpu}{CPU}{Central Processing Unit}
\newacronym{csv}{CSV}{Comma-Separated Values}
\newacronym{ctat}{CTAT}{Complementary To Absolute Temperature}
\newacronym{cuba}{CUBA}{Current Based}
\newacronym{dac}{DAC}{Digital to Analog Converter}
\newacronym{dcn}{DCN}{Dorsal Column Nuclei}
\newacronym{dc}{DC}{Direct Current}
\newacronym{dpi}{DPI}{Differential Pair Integrator}
\newacronym{dpss}{DPSS}{Dendritic Prediction of Somatic Spiking}
\newacronym{drg}{DRG}{Dorsal Root Ganglion}
\newacronym{dr}{DR}{Dual Rail}
\newacronym{dsnn}{D-SNN}{Deep Spiking Neural Network}
\newacronym{dsp}{DSP}{Digital Signal Processing}
\newacronym{ds}{DS}{Delay Sensitive}
\newacronym{dvs}{DVS}{Dynamic Vision Sensor}
\newacronym{eda}{EDA}{Electronic Design Automation}
\newacronym{esn}{ESN}{Echo State Network}
\newacronym{fac}{FAC}{Facilitatory Trace}
\newacronym{fecap}{FeCap}{Ferroelectric Capacitor}
\newacronym{fet}{FET}{Field Effect Transistor}
\newacronym{fft}{FFT}{Fast Fourier Transform}
\newacronym{fifo}{FIFO}{First In First Out Register}
\newacronym{fi}{FI}{Frequency vs. Current}
\newacronym{fm}{FM}{Frequency Modulated}
\newacronym{fo}{FO}{First Order}
\newacronym{fpga}{FPGA}{Field Programmable Gate Array}
\newacronym{fsm}{FSM}{Finite State Machine}
\newacronym{ftj}{FTJ}{Ferroelectric Tunnel Junction}
\newacronym{ft}{FT}{Fourier transform}
\newacronym{gmm}{GMM}{Gaussian mixture models}
\newacronym{gpu}{GPU}{Graphic Process Unit}
\newacronym{hp}{HP}{High Pass Filter}
\newacronym{i2c}{I$^2$C}{Inter Integrated Circuit}
\newacronym{ic}{IC}{Integrated Circuit}
\newacronym{if}{IF}{Integrate and Fire}
\newacronym{imc}{IMC}{In-Memory Computing}
\newacronym{iot}{IoT}{Internet of Things}
\newacronym{io}{I/O}{Input and Output}
\newacronym{ip}{IP}{Intellectual Property Macro}
\newacronym{irh}{IRH}{Instantaneous Rate Histogram}
\newacronym{isi}{ISI}{Interspike Interval}
\newacronym{iwta}{I-WTA}{Inverted Winner-Take-All}
\newacronym{knn}{K-NN}{K-Nearest Neighbors}
\newacronym{lcadc}{LC-ADC}{Level Crossing Analog Digital Converter}
\newacronym{lda}{LDA}{Linear Discriminant Analysis}
\newacronym{lif}{LIF}{Leaky \myglsentry{if}}
\newacronym{llm}{LLM}{Large Language Model}
\newacronym{lpf}{LPF}{Low Pass Filter}
\newacronym{lstm}{LSTM}{Long Short-Term Memory}
\newacronym{ltp}{LTP}{Long Term Plasticity}
\newacronym{lut}{LUT}{Look Up Table}
\newacronym{lvds}{LVDS}{Low Voltage Differential Signalling}
\newacronym{mac}{MAC}{Multiply Accumulate}
\newacronym{mems}{MEMS}{Micro Electro-Mechanical System}
\newacronym{mim}{MIM}{Metal Insulator Metal}
\newacronym{mlm}{MLM}{Multi Layer Mask}
\newacronym{mlp}{MLP}{Multi Layer Perceptron}
\newacronym{ml}{ML}{Machine Learning}
\newacronym{mnist}{MNIST}{modified National Institute of Standards and Technology database}
\newacronym{mom}{MOM}{Metal Oxide Metal}
\newacronym{mos}{MOS}{Metal Oxide Semiconductor}
\newacronym{mtj}{MTJ}{Magnetic Tunnel Junction}
\newacronym{mvm}{MVM}{Matrix-Vector Multiplication}
\newacronym{nas}{NAS}{Neuromorphic Auditory Sensor}
\newacronym{nda}{NDA}{Non Disclosure Agreement}
\newacronym{nmda}{NMDA}{N-methyl-D-aspartate}
\newacronym{nni}{NNI}{Neural Network Intelligence}
\newacronym{nn}{NN}{Neural Network}
\newacronym{noc}{NoC}{Network on Chip}
\newacronym{nvm}{NVM}{Non-Volatile Memory}
\newacronym{oa}{OA}{OpenAccess}
\newacronym{opamp}{OPAMP}{Operational Amplifier}
\newacronym{ota}{OTA}{Operational Transconductance Amplifier}
\newacronym{pca}{PCA}{Principal Component Analysis}
\newacronym{pcb}{PCB}{Printed Circuit Board}
\newacronym{pcfb}{PCFB}{Pre-Charge Full Buffer}
\newacronym{pchb}{PCHB}{Pre-Charge Half Buffer}
\newacronym{pcm}{PCM}{Phase Change Material}
\newacronym{pc}{PC}{Pacini}
\newacronym{pdk}{PDK}{Process Development Kit}
\newacronym{pd}{PD}{Phase Detector}
\newacronym{pll}{PLL}{Phase-Locked Loop}
\newacronym{posfet}{POS-FET}{Piezoelectric Oxide Semiconductor Field Effect Transistor}
\newacronym{prs}{PRS}{Production Rule Set}
\newacronym{pr}{P\&R}{Place and Route}
\newacronym{psc}{PSC}{Post Synaptic Current}
\newacronym{ptat}{PTAT}{Proportional To Absolute Temperature}
\newacronym{pv}{PV}{Parietal Ventral Area}
\newacronym{qdi}{QDI}{Quasi Delay Insensitive}
\newacronym{qfp}{QFP}{Quad-Flat Package}
\newacronym{ra1}{RA}{Rapid-Adapting I}
\newacronym{ra2}{RA2}{Rapid-Adapting II}
\newacronym{rafr}{RaFr}{Radio Frequency}
\newacronym{ram}{RAM}{Random Access Memory}
\newacronym{rbssg}{RBSSG}{Reverse Bitwise Synthetic Spike Generator}
\newacronym{rf}{RF}{Receptive Field}
\newacronym{rl}{RL}{Reinforcement Learning}
\newacronym{rnn}{RNN}{Recurrent Neural Network}
\newacronym{roi}{ROI}{Region Of Interest}
\newacronym{rpe}{RPE}{Reward Prediction Error}
\newacronym{s1}{S1}{Somatosensory Primary Cortex}
\newacronym{s2}{S2}{Somatosensory Secondary Cortex}
\newacronym{sa1}{SA1}{Slow-Adapting I}
\newacronym{sa2}{SA2}{Slow-Adapting II}
\newacronym{sdsp}{SDSP}{Spike-Driven Synaptic Plasticity}
\newacronym{sfd}{SFD}{Spike Frequency Divider}
\newacronym{shf}{S-HF}{Spike-based Hold \& Fire}
\newacronym{sig}{S-IG}{Spike-based Integrate \& Fire}
\newacronym{simd}{SIMD}{Single Instruction, Multiple Data}
\newacronym{slpf}{S-LPF}{Spike-based Low Pass Filter}
\newacronym{smd}{SMD}{Surface Mount Device}
\newacronym{smu}{SMU}{Source Measurement Unit}
\newacronym{soc}{SoC}{System on Chip}
\newacronym{sota}{SOTA}{State-Of-The-Art}
\newacronym{so}{SO}{Second Order}
\newacronym{spice}{SPICE}{Simulation Program with Integrated Circuit Emphasis}
\newacronym{spi}{SPI}{Serial Peripheral Interface}
\newacronym{spll}{sPLL}{Spiking Phase-Locked Loop}
\newacronym{src}{SRC}{Sparse Representation Classifier}
\newacronym{srdp}{SRDP}{Spike-Rate-Dependent Plasticity}
\newacronym{srm}{SRM}{Simple Response Model}
\newacronym{stdp}{STDP}{Spike Timing Dependent Plasticity}
\newacronym{stp}{STP}{Short Term Plasticity}
\newacronym{svm}{SVM}{Support Vector Machine}
\newacronym{syn}{SYN}{Synaptic Trace}
\newacronym{tc}{TC}{Temporal Contrast}
\newacronym{tde}{TDE}{Time Difference Encoder}
\newacronym{tpu}{TPU}{Tensor Processing Unit}
\newacronym{trg}{TRG}{Trigger Trace}
\newacronym{ttfs}{TTFS}{Time to first spike}
\newacronym{uc}{$\mu$C}{Microcontroller}
\newacronym{vco}{VCO}{Voltage Controlled Oscillator}
\newacronym{vpd}{VP-d}{Victor-Purpura Distance}
\newacronym{wta}{WTA}{Winner Take All}

\newacronym{alif}{ALIF}{Adaptive \myglsentry{lif}}
\newacronym{ann}{ANN}{Artificial \myglsentry{nn}}
\newacronym{asic}{ASIC}{Application Specific \myglsentry{ic}}
\newacronym{cmos}{CMOS}{Complementary \myglsentry{mos}}
\newacronym{cnn}{CNN}{Convolutional \myglsentry{nn}}
\newacronym{cstdp}{C-STDP}{Calcium \myglsentry{stdp}}
\newacronym{ddpi}{DDPI}{Double \myglsentry{dpi}}
\newacronym{dnn}{DNN}{Deep \myglsentry{nn}}
\newacronym{eai}{Edge-AI}{Edge \myglsentry{ai}}
\newacronym{enn}{ENN}{event-based \myglsentry{nn}}
\newacronym{epsc}{EPSC}{Excitatory \myglsentry{psc}}
\newacronym{exlif}{ExLIF}{Exponential \myglsentry{lif}}
\newacronym{fefet}{FeFET}{Ferroelectric \myglsentry{fet}}
\newacronym{gpgpu}{GPGPU}{General-Purpose computing on \myglsentry{gpu}}
\newacronym{ipsc}{IPSC}{Inhibitory \myglsentry{psc}}
\newacronym{moscap}{MOSCAP}{\myglsentry{mos} Capacitor}
\newacronym{mosfet}{MOSFET}{\myglsentry{mos} \myglsentry{fet}}
\newacronym{nmnist}{N-MNIST}{Neuromorphic \myglsentry{mnist}}
\newacronym{nmos}{nMOS}{n-type \myglsentry{mos}}
\newacronym{oxram}{OxRAM}{Oxide-based resistive \myglsentry{ram}}
\newacronym{pka}{PKA}{\myglsentry{camp}-depended Protein Kinase}
\newacronym{pmos}{pMOS}{p-type \myglsentry{mos}}
\newacronym{rram}{RRAM}{Resistive \myglsentry{ram}}
\newacronym{sadc}{sADC}{spiking \myglsentry{adc}}
\newacronym{scnn}{sCNN}{spiking \myglsentry{cnn}}
\newacronym{snn}{SNN}{Spiking \myglsentry{nn}}
\newacronym{sodpi}{SoDPI}{Second-order \myglsentry{dpi}}
\newacronym{sram}{SRAM}{Static \myglsentry{ram}}
\newacronym{tstdp}{T-STDP}{Triplet \myglsentry{stdp}}

\newacronym{adexlif}{AdExLIF}{Adaptive \myglsentry{exlif}}
\newacronym{nfet}{n-FET}{\myglsentry{nmos} \myglsentry{fet}}
\newacronym{pfet}{p-FET}{\myglsentry{pmos} \myglsentry{fet}}

\newacronym{sma}{SMA}{SubMiniature version A}
\newacronym{wrota}{WR-OTA}{Wide Range \myglsentry{ota}}
\newacronym{sstdp}{S-STDP}{Stochastic \myglsentry{stdp}}
\newacronym{neuop}{NeuOp}{Neuron spike Operation}
\newacronym{synop}{SynOp}{Synaptic Operation}
\newacronym{fdsoi}{FDSOI}{Fully Depleted Silicon On Insulator}
\newacronym{qif}{QIF}{Quadratic \myglsentry{if}}
\newacronym{ff}{FF}{edge triggered Flip-Flop}
\newacronym{drtp}{DRTP}{Direct Random Target Propagation}
\newacronym{etlp}{ETLP}{??}
\newacronym[longplural={random hidden layer neural networks also referred to as Extreme Learning Machines}]{elm}{ELM}{random hidden layer neural network also referred to as Extreme Learning Machine}
\newacronym{nef}{NEF}{Neural Engineering Framework}
\newacronym{si}{Si}{Silicon}
\newacronym{ldi}{LDI}{Log Domain Integrator}
\newacronym{px}{PX}{Pulse Extender}
\newacronym{ode}{ODE}{Ordinary Differential Equation}
\newacronym{agc}{AGC}{Automatic Gain Control}
\newacronym{psp}{PSP}{Post Synaptic Potential}
\newacronym{dut}{DUT}{Design Under Test}
\newacronym{lrs}{LRS}{Low Resistive State}
\newacronym{hrs}{HRS}{High Resistive State}
\newacronym{cimc}{CiMC}{Compute in Memory Controller}
\newacronym{mwp}{MWP}{Multi Project Wafer}
\newacronym{vlsi}{VLSI}{Very Large Scale Integration}

\makeglossaries

\usepackage[giveninits=true,minbibnames=1,maxbibnames=6,url=false,mincitenames=1,maxcitenames=2,
backend=biber,sorting=none,citestyle=numeric-comp,bibstyle=ieee]{biblatex}

\AtBeginBibliography{\footnotesize}
\AtEveryBibitem{
    \clearfield{urlyear}
    \clearfield{urlmonth}
    \clearfield{urlday}
    \clearlist{language}
    \clearfield{origlanguage}
    \clearfield{url}
    \clearfield{issn}
    \clearfield{isbn}
    \clearfield{number}
    \clearfield{note}
    \clearlist{location}
    \clearfield{month}
    \clearname{editor}
    \clearlist{publisher}
    \clearfield{address}
    \clearfield{series}
}

\let\oldcite\cite
\renewcommand{\cite}[1]{{\mbox{\oldcite{#1}}}}

\usepackage{doi}

\begin{document}

\title{A scalable event-driven spatiotemporal \\ feature extraction circuit

\thanks{
We thank Ton Juny Pina and Philipp Klein for their help with PCB soldering, and Thorben Schoepe for the chip photography.
The work was funded by grants: EU H2020: NeuTouch (813713), BeFerroSynaptic (871737) and MANIC (861153); DFG: memTDE (441959088 - SPP 2262 MemrisTec 422738993) and  NMVAC (432009531).
We acknowledge the financial support of the CogniGron research center and the Ubbo Emmius Funds (University of Groningen).
CRedit:
Conceptualisation: H. G., M. M.; Methodology: E. C., H. G., M. M.; Software/Hardware: M. C., H. G., E. J., M. M., O. R., W. S. G.; Investigation:  H. G., M. M.; Writing: E.C., M. C., H. G., M. M., O. R.; Visualisation: M. C., H. G., M. M., O. R.;  Supervision: E. C.\\
Affiliations: \\
\textsuperscript{1} Bio-Inspired Circuits and Systems (BICS) Lab, Zernike Institute for Advanced Materials, University of Groningen, Netherlands\newline
\textsuperscript{2} Groningen Cognitive Systems and Materials Center (CogniGron), University of Groningen, Netherlands\newline
\textsuperscript{3} Micro- and Nanoelectronic Systems (MNES), Technische Universit\"at Ilmenau, Germany \newline
\textsuperscript{4} Istituto Italiano di Tecnologia, Genova, Italy\\
\textsuperscript{*} Equal contribution \\ Corresponding authors: \{h.r.greatorex, m.mastella\}@rug.nl
}

}

\author{\IEEEauthorblockA{\textbf{Hugh Greatorex}\textsuperscript{1,2,*}, \textbf{Michele Mastella}\textsuperscript{1,2,*}, \textbf{Ole Richter}\textsuperscript{1,2}, \textbf{Madison Cotteret}\textsuperscript{1,2,3}, \\ \textbf{Willian Soares Girão}\textsuperscript{1,2}, \textbf{Ella Janotte}\textsuperscript{1,2,4},  \textbf{Elisabetta Chicca}\textsuperscript{1,2}}
}

\maketitle

\begin{abstract}
Event-driven sensors, which produce data only when there is a change in the input signal, are increasingly used in applications that require low-latency and low-power real-time sensing, such as robotics and edge devices.
To fully achieve the latency and power advantages on offer however, similarly event-driven data processing methods are required.
A promising solution is the \gls{tde}: an event-based processing element which encodes the time difference between events on different channels into an output event stream. 
In this work we introduce a novel \gls{tde} implementation on CMOS. 
The circuit is robust to device mismatch and allows the linear integration of input events. 
This is crucial for enabling a high-density implementation of many \glspl{tde} on the same die, and for realising real-time parallel processing of the high-event-rate data produced by event-driven sensors.

\end{abstract}


\section{Introduction}
The emergence of event-based sensors has opened up new possibilities for applications that demand low-latency and energy-efficient data processing.
Unlike traditional frame-based sensors that capture information at fixed intervals, event-based sensors transmit data only when a change is detected. 
This makes them highly suitable for real-time decision-making tasks across various sensory modalities, such as vision, audio, and tactile sensing~\cite{delbruck2010activity, bartolozzi2020, lyon1988}. 
These sensors not only offer low latency but also enable high temporal resolution and low power consumption, provided the event streams remain sparse.
As a result, there has been growing interest in developing event-based processing units and circuits that can handle these streams while retaining the inherent benefits of the event-driven paradigm~\cite{Yao_etal24,liu_event-driven_2019,afshar_event-based_2020}.

In this paper, we present a novel circuit that implements an improved version of the event-based processing element known in literature as the \gls{tde}~\cite{schoepe_finding_2024, Schoepe_etal24b,Schoepe_2023,mastella_hardware_2021, dangelo20_doi, chiavazza_2023_CVPR}.
While there has been a growing propensity to design event-based circuits that operate in the digital domain~\cite{Davies_etal18, Merolla_etal14_aux}, the original breakthroughs in this area were achieved through analog implementations that leverage the underlying physics of the devices~\cite{Mead89_doi, mahowald1991silicon}. 
Our design stays true to this analog approach, using transistors operating in subthreshold with currents in the range of \mbox{\qty{}{\pA}}.  
The improved circuit in this work is designed to overcome issues related to mismatch, making it more scalable for large arrays.
We demonstrate the circuit's application to an event-vision task, highlighting its ability to extract spatiotemporal patterns from event-based data ~\cite{gallego2020event}. 
This makes it a strong candidate for edge applications, where low energy consumption and minimal latency are crucial for real-time sensory data processing.

\begin{figure}[!htb]
    \centering
    \includegraphics[width=0.8\linewidth]{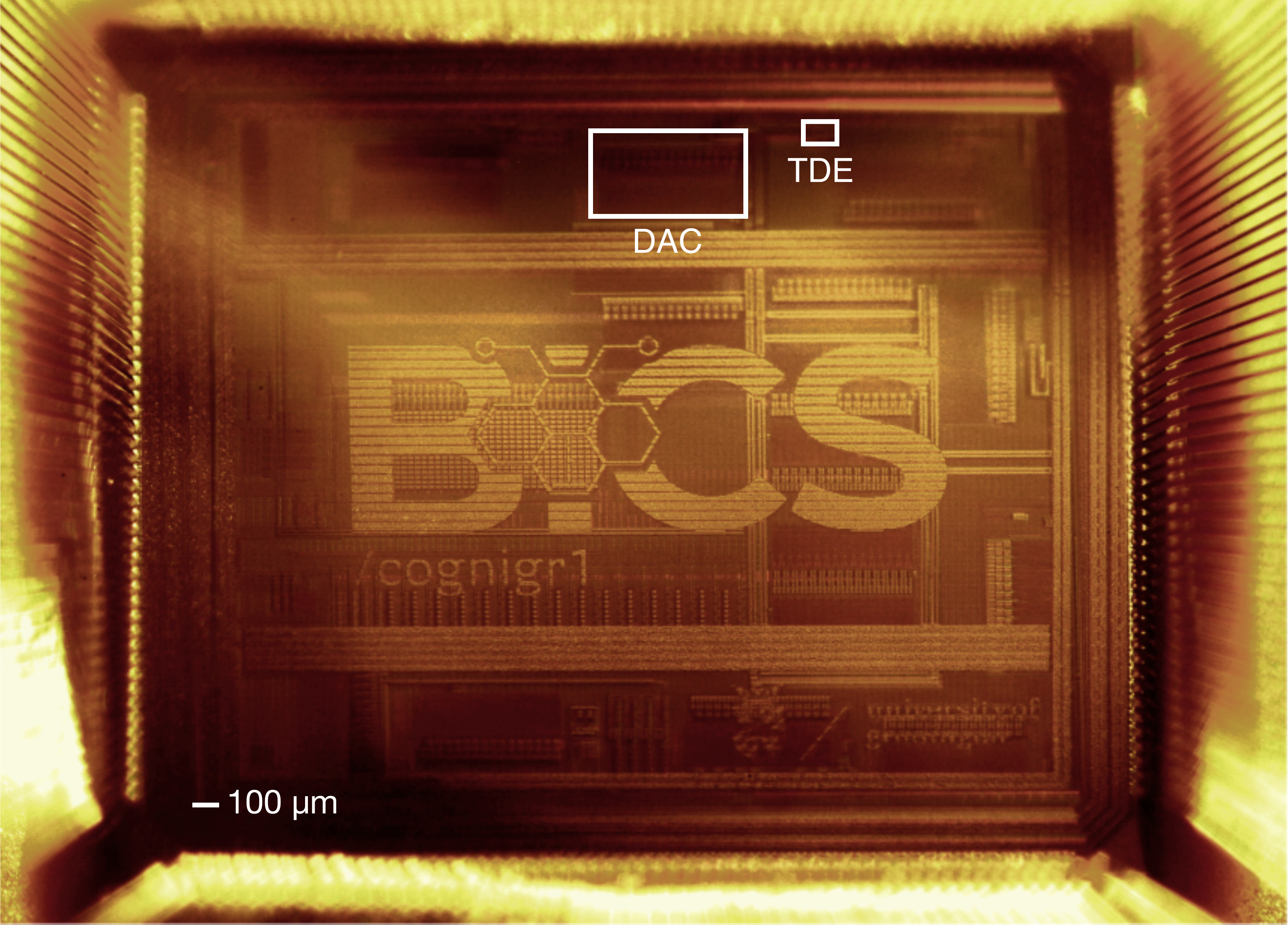}
    \caption{Photograph of the realised ``cognigr1'' \gls{asic}, fabricated in the XFAB \qty{180}{\nm} technology. The boxes highlight the location of the structures on the die. The total size of the \gls{tde} circuit is \mbox{\qty{19}{\um}$\times$\qty{56}{\um}} including guard rings.}
    \label{fig:chip}
\end{figure}

\section{Methods}

\begin{figure*}[!htb]
    \centering
    \includegraphics[width=\linewidth]{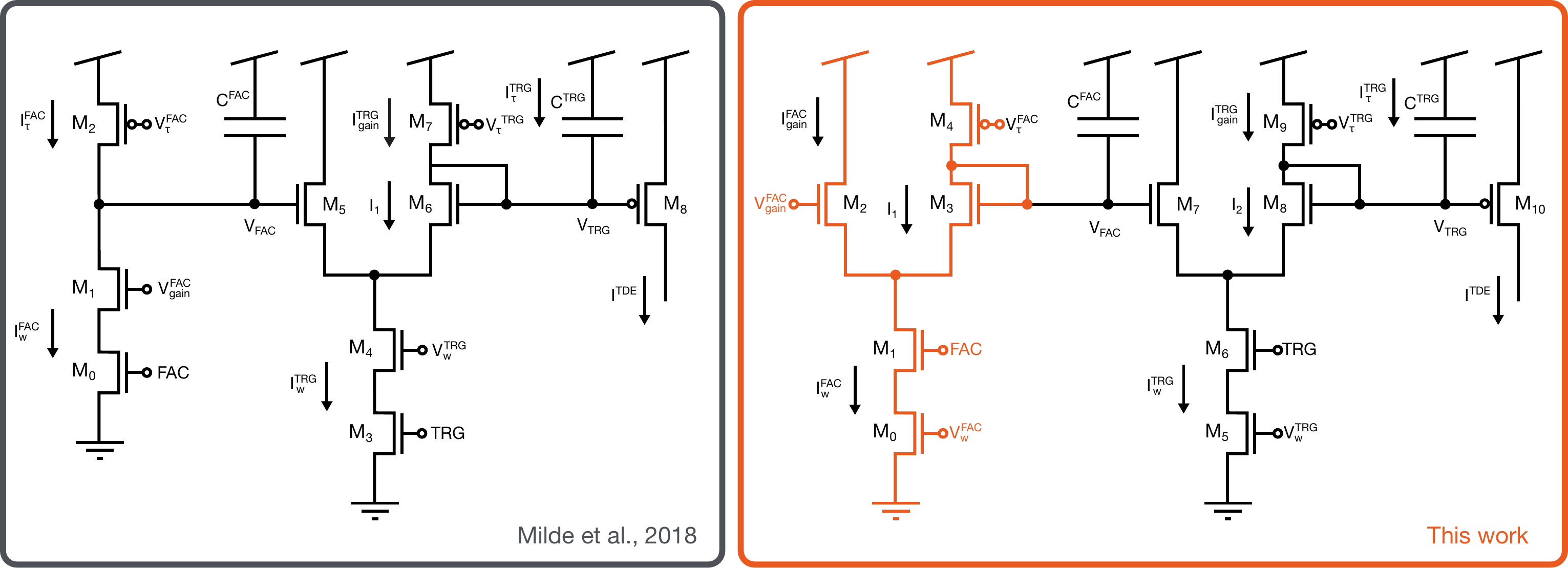}
    \caption{The circuit schematics of the \gls{tde} synapse. The circuit initially proposed in \textcite{milde18_doi} consists of a single discharge branch in the facilitatory block and a \gls{dpi}~\cite{Bartolozzi_Indiveri07b} in the trigger block. 
    The improved circuit presented in this work makes use of a \gls{dpi} circuit in both blocks to ensure linear integration of events. Modifications with respect to the old circuit are highlighted.}
    \label{fig:circuits}
\end{figure*}

\subsection{Improved TDE Circuit}

The circuit proposed in this work (Fig.~\ref{fig:circuits}) is compared to the \gls{tde} circuit originally proposed by~\mbox{\textcite{milde18_doi}}, following previous implementations of CMOS motion detectors~\cite{kramer97_doi, indiveri_1999}. 
The \gls{tde} circuit generates an exponentially decaying current with an initial magnitude proportional to the negative exponential of the time difference between two spikes~\cite{greatorex_etal2024}. 
This current is fed into a \gls{lif} neuron~\cite{livi_current-mode_2009} to produce output spikes with an instantaneous frequency with the same proportionality to the input time difference. 
This is achieved by the two distinct ``facilitatory'' and ``trigger'' integrator blocks in the \gls{tde} circuit, which we refer to collectively as the \gls{tde} synapse. The inputs to each block are the two input channels of the \gls{tde}.

\begin{figure}[!htb]
    \centering
    \includegraphics[width=0.9\linewidth]{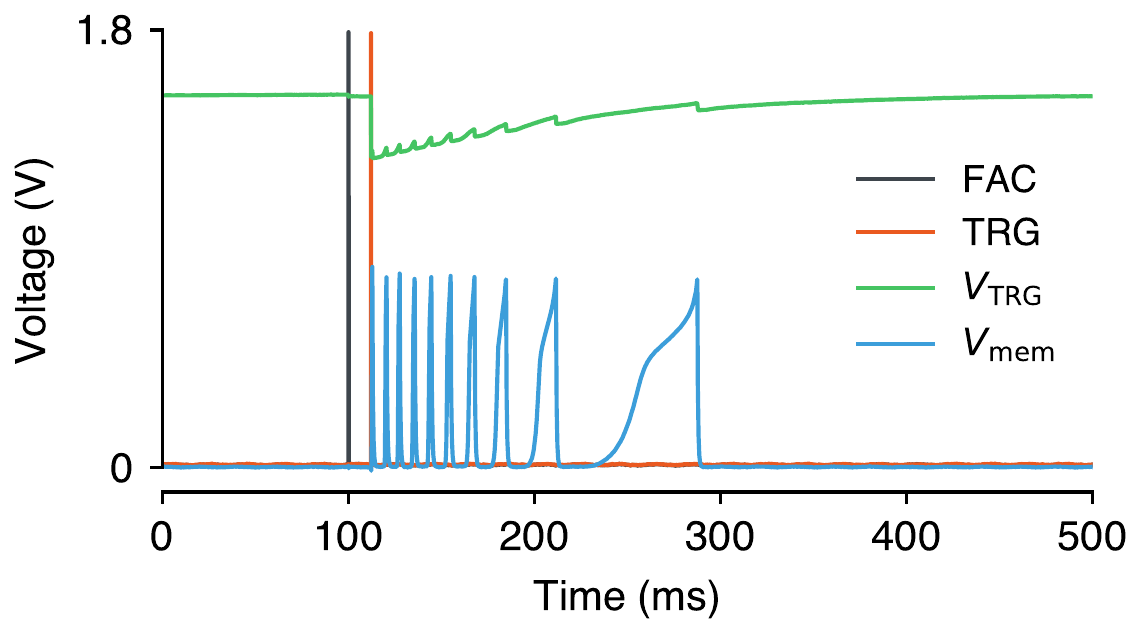}
    \caption{
    Silicon measurements of the \gls{tde} circuit. The \gls{tde} received two sequential FAC and TRG input events with a time difference of \mbox{\qty{12}{ms}}. 
    The response of the trigger block, $V_{\text{TRG}}$, generates an \gls{epsc} through transistor $\text{M}_{10}$ which is integrated by the neuron, eliciting a spiking response observed through the membrane potential $V_{\text{mem}}$.}
    \label{fig:circuit_measurement}
\end{figure}

Both blocks include a time-decaying voltage trace initiated by a digital pulse (or event). 
When an event arrives at the facilitatory input (FAC), a decaying voltage trace is initiated. 
When an event arrives at the trigger input (TRG), a synaptic current with amplitude proportional to the immediate value of the trigger trace (relative to the power supply) is generated. 
This current is integrated by a neuron circuit. 
The closer the facilitatory and trigger spikes are in time, the higher the amplitude of the facilitatory trace at the arrival of the TRG spike and of the consequent \gls{epsc}. 
Therefore, the resulting output spikes encode the input time difference both in the number of output spikes and their \gls{isi}, or equivalently their instantaneous firing rate.
In this way the \gls{tde} circuit acts as an asymmetric correlation detector on the two inputs and encodes the time difference between input events in an analog fashion. 

\subsection{Testing setup}
\label{subsec:testing_setup}
The cognigr1 \gls{ic}~\cite{mastella_23, cotteret_23, richter_23} was fabricated in CMOS (Fig.~\ref{fig:chip}) using a \qty{180}{\nm} technology node.
The biases of the circuit are provided using a subthreshold \gls{dac} present on the die. 
All tests were carried out using a custom setup with oscilloscopes and a \mbox{microcontroller} with a Python interface supplying configuration information and events. 

\subsection{Optical flow task}

\begin{figure*}[!htb]
    \centering
    \begin{subfigure}[b]{0.74\linewidth}
        \centering
        \includegraphics[width=\linewidth]{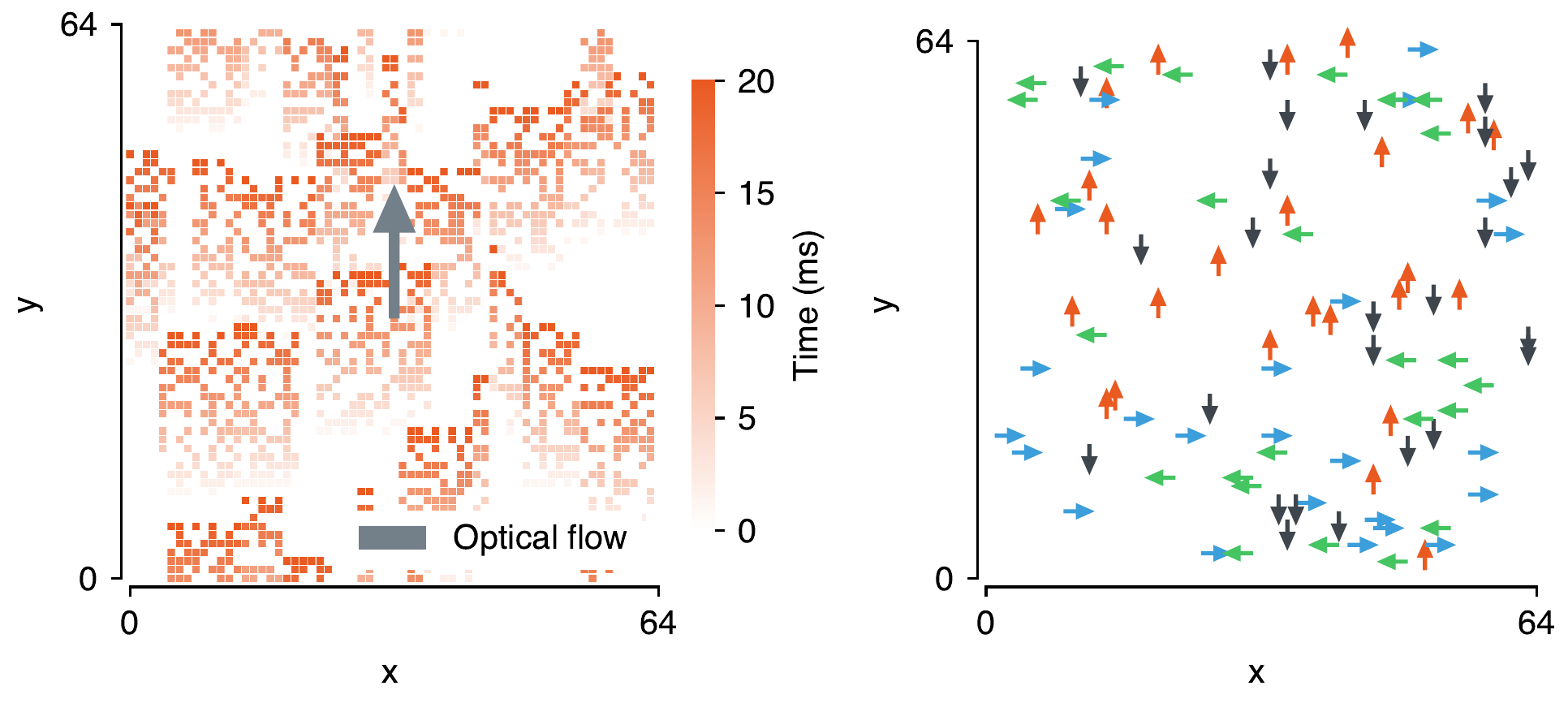}
    \end{subfigure}
    \hfill
    \begin{subfigure}[b]{0.25\linewidth}
        \centering
        \includegraphics[width=\linewidth]{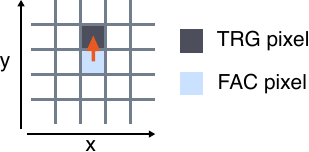}
        \vspace{50pt}
    \end{subfigure}
    \caption{Left: Simulated event-based camera data of a textured surface moving in a vertical direction. 
    Center: The x-y space was randomly and sparsely sampled by 100 \glspl{tde} with an equal ratio of cardinal orientations. 
    Right: Arrows represent the connectivity of FAC and TRG circuit inputs to neighbouring event-based camera pixels.}
    \label{fig:optical_flow_task}
\end{figure*}

The \gls{tde} circuit was applied to an event-based vision task. 
We simulated event-based camera~\cite{delbruck2010activity} data (Fig.~\ref{fig:optical_flow_task}) to evaluate the circuit's capability to detect optical flow.
A synthetic textured surface consisting of multiple irregularities and patterns moving upwards was used to generate a stream of x-y address events. 
Each \gls{tde} in the network was configured to receive the events from two particular pixels on its FAC and TRG inputs. 
In this way, each \gls{tde} was configured to have a certain direction sensitivity, determined by its 2-pixel receptive field. 
One hundred \glspl{tde} were randomly distributed across the visual field, with equal proportions oriented in the four cardinal directions. 
Since the cognigr1 \gls{ic} hosts only a single \gls{tde} circuit, the event-based camera data was sequentially applied to its input.
If multiple \gls{tde} circuits are present in an on-chip array, further analysis would be needed to examine the impact of mismatch on task performance.

\section{Results and Discussion}

\begin{figure}[!htb]
    \centering
    \includegraphics[width=0.9\linewidth]{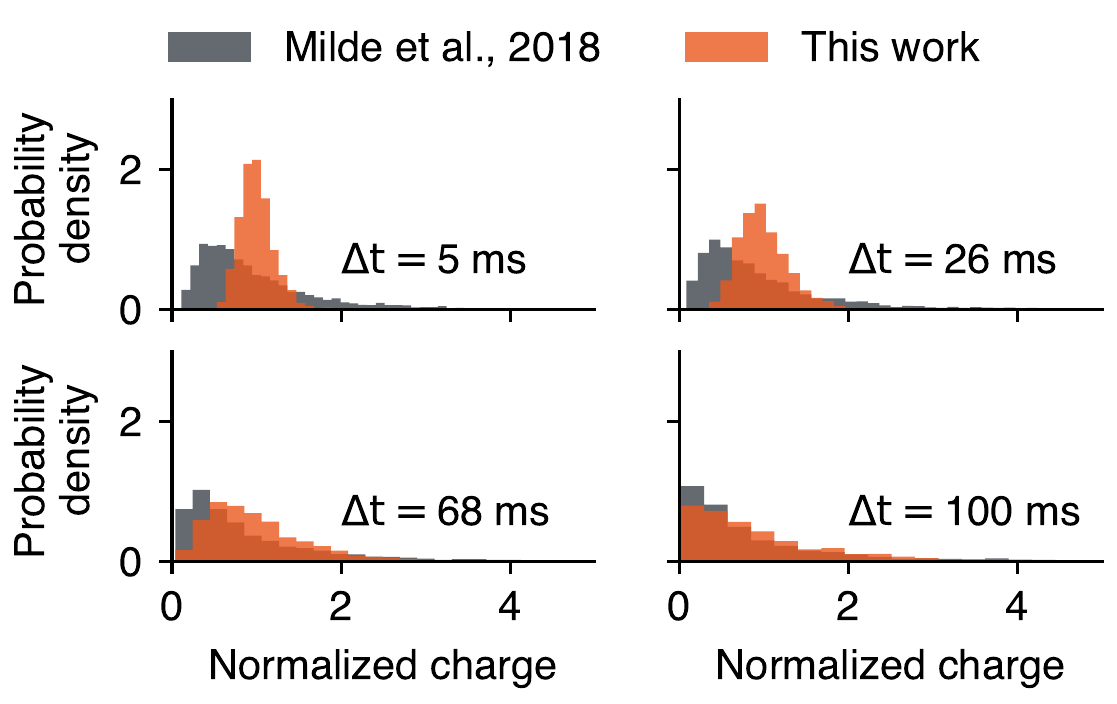}
    \caption{
    Monte Carlo simulation results comparing our circuit with~\textcite{milde18_doi}. 
    For a given time difference, $\Delta t$, the charge transmitted by the \gls{tde} synapse was measured for 2000 simulated instances of the circuit. 
    This charge was then normalized relative to the average charge for a given $\Delta t$. }
    \label{fig:monte_carlo}
\end{figure}

\subsection{Circuit measurements}

Figure~\ref{fig:circuit_measurement} shows measurements of input and output signals of the \gls{tde} circuit.
Digital pulses, representing events or spikes, were applied to the FAC and TRG transistors of the circuit ($\text{M}_{1}$ and $\text{M}_{6}$ respectively). 
The time difference between the two pulses is encoded by $V_\text{TRG}$ which generates an \gls{epsc} with an amplitude that encodes the time between input events. 
This current is then fed into the neuron circuit eliciting a spiking response that can be observed through the membrane potential $V_{\text{mem}}$.

\subsection{Monte Carlo analysis}

\begin{figure}[!htb]
    \centering
    \includegraphics[width=0.85\linewidth]{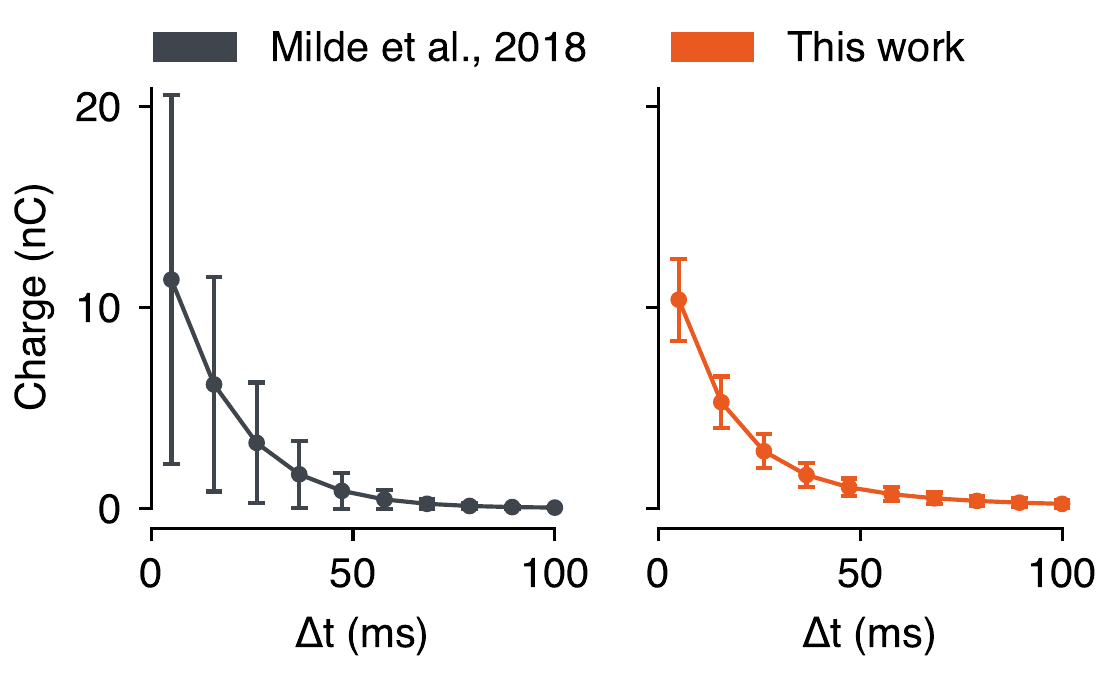}
    \caption{The average charge transmitted plotted against $\Delta t$, error bars show the standard deviation across Monte Carlo simulations.}
    \label{fig:monte_carlo_condensed}
\end{figure}

To validate the improvement of our circuit with respect to the circuit proposed in~\textcite{milde18_doi}, Monte Carlo simulations were performed and analysed using the Cadence Spectre tool. We simulated 
2000 different instances of both circuits over a range of $\Delta t$ values, sampling from a realistic distribution of physical transistor variations. 
Figure~\ref{fig:monte_carlo} shows the distribution of charge sourced by both circuits, calculated by integrating the current $I^{\text{TDE}}$ (see Fig.~\ref{fig:circuits}).  
Figure~\ref{fig:monte_carlo_condensed} shows the mean and standard deviation of each Monte Carlo simulation, demonstrating the improved robustness of the newly proposed circuit to device mismatch. 
We found an average reduction in the coefficient of variation of 61\% averaged across $\Delta t$.

\subsection{Optical flow task}

\begin{figure}[!htb]
    \centering
    \includegraphics[width=0.9\linewidth]{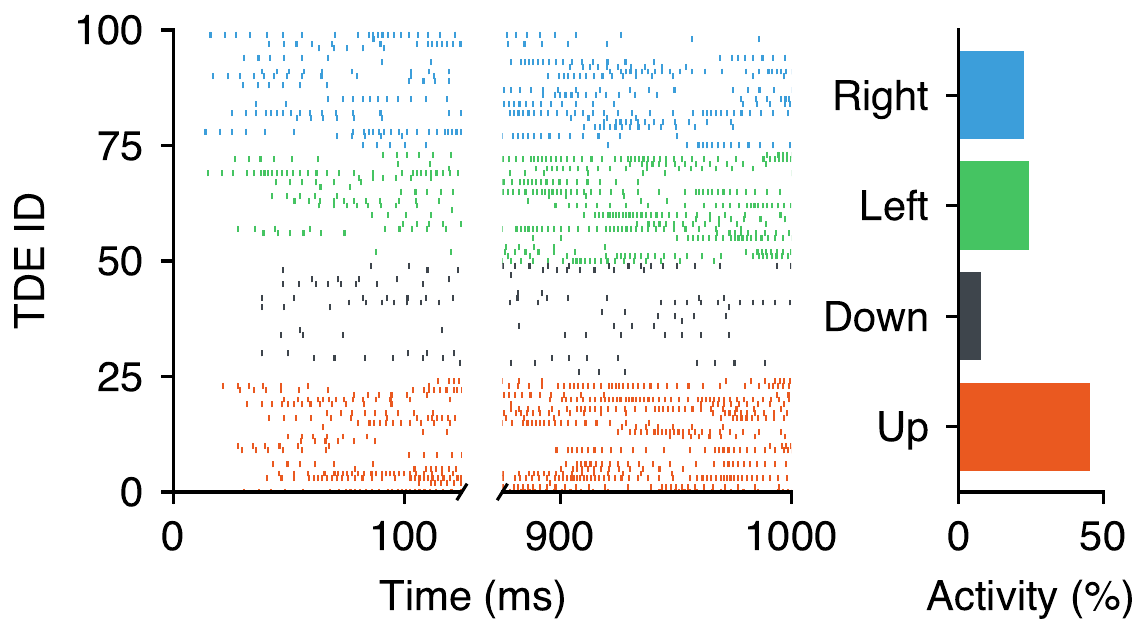}
    \caption{Left: A raster plot of the measured spiking responses of the \glspl{tde} in the network. 
    Right: The measured number of spikes for the four \gls{tde} orientations is expressed as a fraction of the total network activity during the event data duration.}
    \label{fig:optical_flow_task_results}
\end{figure}

Figure~\ref{fig:optical_flow_task_results} shows the spiking response of the 100 \gls{tde} units in the network run on the cognigr1 \gls{ic}. 
The response of each orientation as a percentage of the total network activity shows a significant preference towards the direction of optical flow. 
The \glspl{tde} oriented perpendicular to the direction of optical flow (left and right) exhibit a greater response in comparison to those orientanted in the null direction (down). 
This can be explained by considering that an ideal sensor would produce exactly coincident spikes from pixels aligned along a moving edge.

Instead, the inherent asynchrony of a physical \gls{aer}~\cite{Boahen1998} sensor produces some jitter which was also introduced into the synthetic data. 
This results in very small time differences at the input of a \gls{tde} with a receptive field aligned to a moving edge, leading to high activation of \glspl{tde} oriented perpendicular to the direction of motion.
However, when sampled across the entire visual field, this effect averages out, resulting in roughly equal responses in the orthogonal directions.

\section{Conclusion}

In this work, we designed and silicon-verified an enhanced version of the \gls{tde} event-based computational primitive. 
The modifications we made to the circuit significantly improved its robustness to device mismatch, as confirmed by comprehensive Monte Carlo simulations. 
These simulations demonstrated the circuit's suitability for integration into high-density array structures on CMOS technology, thereby paving the way for efficient scaling in practical applications.
To explore the circuit's potential for highly parallel real-time processing, we tested it on an optical flow detection task utilizing event-based synthetic data derived from a moving texture. 
This application highlighted the circuit's capabilities not only in handling real-time data but also in its efficiency in processing the information generated by event-driven sensors. 



\printbibliography

@Inbook{bartolozzi2020,
author="Bartolozzi, Chiara
and Glover, Arren
and Donati, Elisa",
editor="Thakor, Nitish V.",
title="Neuromorphic Sensing, Perception and Control for Robotics",
bookTitle="Handbook of Neuroengineering",
year="2020",
publisher="Springer Singapore",
pages="1--31",
isbn="978-981-15-2848-4",
doi="10.1007/978-981-15-2848-4_116-1",
url="https://doi.org/10.1007/978-981-15-2848-4_116-1"
}

@article{schoepe_finding_2024,
	title = {Finding the gap: neuromorphic motion-vision in dense environments},
	volume = {15},
	copyright = {2024 The Author(s)},
	issn = {2041-1723},
	shorttitle = {Finding the gap},
	url = {https://www.nature.com/articles/s41467-024-45063-y},
	doi = {10.1038/s41467-024-45063-y},
	language = {en},
	number = {1},
	urldate = {2024-07-09},
	journal = {Nature Communications},
	author = {Schoepe, Thorben and Janotte, Ella and Milde, Moritz B. and Bertrand, Olivier J. N. and Egelhaaf, Martin and Chicca, Elisabetta},
	month = jan,
	year = {2024},
	note = {Publisher: Nature Publishing Group},
	pages = {817},
}

@INPROCEEDINGS{Schoepe_etal24b,
  author={Schoepe, Thorben and Drix, Damien and Schüffny, Franz Marcus and Miko, Rebecca and Sutton, Samuel and Chicca, Elisabetta and Schmuker, Michael},
  booktitle={IEEE ISCAS 2024}, 
  title={Odour Localization in Neuromorphic Systems}, 
  year={2024},
  volume={},
  number={},
  pages={1-5},
  doi={10.1109/ISCAS58744.2024.10558186}}

@article{Schoepe_2023,
doi = {10.1088/2634-4386/acdaba},
url = {https://dx.doi.org/10.1088/2634-4386/acdaba},
year = {2023},
month = {jun},
publisher = {IOP Publishing},
volume = {3},
number = {2},
pages = {024009},
author = {Thorben Schoepe and Daniel Gutierrez-Galan and Juan P Dominguez-Morales and Hugh Greatorex and Angel Jimenez-Fernandez and Alejandro Linares-Barranco and Elisabetta Chicca},
title = {Closed-loop sound source localization in neuromorphic systems},
journal = {Neuromorphic Computing and Engineering }
}

@article{liu_event-driven_2019,
	title = {Event-driven sensing for efficient perception: {Vision} and audition algorithms},
	volume = {36},
	shorttitle = {Event-driven sensing for efficient perception},
	number = {6},
	journal = {IEEE Signal Processing Magazine},
	author = {Liu, Shih-Chii and Rueckauer, Bodo and Ceolini, Enea and Huber, Adrian and Delbruck, Tobi},
	year = {2019},
	note = {Publisher: IEEE},
	pages = {29--37},
    doi = {10.1109/MSP.2019.2928127}
}

@article{afshar_event-based_2020,
	title = {Event-{Based} {Object} {Detection} and {Tracking} for {Space} {Situational} {Awareness}},
	volume = {20},
	issn = {1558-1748},
	doi = {10.1109/JSEN.2020.3009687},
	number = {24},
	journal = {IEEE Sensors Journal},
	author = {Afshar, Saeed and Nicholson, Andrew Peter and van Schaik, Andre and Cohen, Gregory},
	month = dec,
	year = {2020},
	note = {Conference Name: IEEE Sensors Journal},
	pages = {15117--15132},
}

@INPROCEEDINGS{livi_current-mode_2009,
  author={Livi, Paolo and Indiveri, Giacomo},
  booktitle={IEEE ISCAS 2009}, 
  title={A current-mode conductance-based silicon neuron for address-event neuromorphic systems}, 
  year={2009},
  volume={},
  number={},
  pages={2898-2901},
  doi={10.1109/ISCAS.2009.5118408}}

@article{Merolla_etal14_aux,
	title = {A million spiking-neuron integrated circuit with a scalable communication network and interface},
	volume = {345},
	issn = {0036-8075},
	doi = {10.1126/science.1254642},
	number = {6197},
	journal = {Science},
	author = {Merolla, P. A. and Arthur, J. V. and Alvarez-Icaza, R. and Cassidy, A. S. and Sawada, J. and Akopyan, F. and Jackson, B. L. and Imam, N. and Guo, C. and Nakamura, Y. and Brezzo, B. and Vo, I. and Esser, S. K. and Appuswamy, R. and Taba, B. and Amir, A. and Flickner, M. D. and Risk, W. P. and Manohar, R. and Modha, D. S.},
	year = {2014},
	note = {Publisher: American Association for the Advancement of Science},
	pages = {668--673},
}

@Inbook{Boahen1998,
author="Boahen, Kwabena A.",
editor="Lande, Tor Sverre",
title="Communicating Neuronal Ensembles between Neuromorphic Chips",
bookTitle="Neuromorphic Systems Engineering: Neural Networks in Silicon",
year="1998",
publisher="Springer US",
address="Boston, MA",
pages="229--259",
isbn="978-0-585-28001-1",
doi="10.1007/978-0-585-28001-1_11",
url="https://doi.org/10.1007/978-0-585-28001-1_11"
}

@inproceedings{delbruck2010activity,
  title={Activity-driven, event-based vision sensors},
  author={Delbr{\"u}ck, Tobi and Linares-Barranco, Bernabe and Culurciello, Eugenio and Posch, Christoph},
  booktitle={IEEE ISCAS 2010},
  pages={2426--2429},
  year={2010},
doi={10.1109/ISCAS.2010.5537149}
}

@inproceedings{mastella_hardware_2021,
	address = {Daegu, Korea},
	title = {A {hardware}-{friendly} {neuromorphic} {spiking} {neural} {network} for {frequency} {detection} and {fine} {texture} {decoding}},
	isbn = {978-1-72819-201-7},
	url = {https://ieeexplore.ieee.org/document/9401377/},
	doi = {10.1109/ISCAS51556.2021.9401377},
	urldate = {2024-07-06},
	booktitle = {IEEE ISCAS 2021},
	author = {Mastella, Michele and Chicca, Elisabetta},
	month = may,
	year = {2021},
	pages = {1--5},
}

@inproceedings{mastella_23, author = {Mastella, Michele and Greatorex, Hugh and Cotteret, Madison and Janotte, Ella and Soares Girao, Willian and Richter, Ole and Chicca, Elisabetta}, title = {Synaptic Normalisation for On-Chip Learning in Analog CMOS Spiking Neural Networks}, year = {2023}, isbn = {9798400701757}, publisher = {Association for Computing Machinery}, address = {New York, NY, USA}, url = {https://doi.org/10.1145/3589737.3606007}, doi = {10.1145/3589737.3606007}, booktitle = {Proceedings of the 2023 International Conference on Neuromorphic Systems}, articleno = {34}, numpages = {4}, location = {Santa Fe, NM, USA}, series = {ICONS '23} }

@INPROCEEDINGS{cotteret_23,
  author={Cotteret, Madison and Richter, Ole and Mastella, Michele and Greatorex, Hugh and Janotte, Ella and Girão, Willian Soares and Ziegler, Martin and Chicca, Elisabetta},
  booktitle={2023 IEEE International Symposium on Circuits and Systems (ISCAS)}, 
  title={Robust Spiking Attractor Networks with a Hard Winner-Take-All Neuron Circuit}, 
  year={2023},
  volume={},
  number={},
  pages={1-5},
  doi={10.1109/ISCAS46773.2023.10181513}}

@inproceedings{richter_23, author = {Richter, Ole and Greatorex, Hugh and Hucko, Benjamin and Cotteret, Madison and Soares Girao, Willian and Janotte, Ella and Mastella, Michele and Chicca, Elisabetta}, title = {A Subthreshold Second-Order Integration Circuit for Versatile Synaptic Alpha Kernel and Trace Generation}, year = {2023}, isbn = {9798400701757}, publisher = {Association for Computing Machinery}, address = {New York, NY, USA}, url = {https://doi.org/10.1145/3589737.3606008}, doi = {10.1145/3589737.3606008}, booktitle = {Proceedings of the 2023 International Conference on Neuromorphic Systems}, articleno = {33}, numpages = {4}, location = {Santa Fe, NM, USA}, series = {ICONS '23} }

@article{mahowald1991silicon,
  title={A silicon neuron},
  author={Mahowald, Misha and Douglas, Rodney},
  journal={Nature},
  volume={354},
  number={6354},
  pages={515--518},
  year={1991},
  publisher={Nature Publishing Group UK London},
url={https://doi.org/10.1038/354515a0},
doi={10.1038/354515a0}
}

@ARTICLE{dangelo20_doi,

AUTHOR={D'Angelo, Giulia  and Janotte, Ella  and Schoepe, Thorben  and O'Keeffe, James  and Milde, Moritz B.  and Chicca, Elisabetta  and Bartolozzi, Chiara },

TITLE={Event-Based Eccentric Motion Detection Exploiting Time Difference Encoding},

JOURNAL={Frontiers in Neuroscience},

VOLUME={14},

YEAR={2020},

URL={https://www.frontiersin.org/journals/neuroscience/articles/10.3389/fnins.2020.00451},

DOI={10.3389/fnins.2020.00451},

ISSN={1662-453X}}

@ARTICLE{kramer97_doi,
  author={Krammer, J. and Koch, C.},
  journal={IEEE Transactions on Circuits and Systems II: Analog and Digital Signal Processing}, 
  title={Pulse-based analog {VLSI} velocity sensors}, 
  year={1997},
  volume={44},
  number={2},
  pages={86-101},
  doi={10.1109/82.554431}}

@article{milde18_doi,
    author = {Milde, M. B. and Bertrand, O. J. N. and Ramachandran, H. and Egelhaaf, M. and Chicca, E.},
    title = "{Spiking Elementary Motion Detector in Neuromorphic
                    Systems}",
    journal = {Neural Computation},
    volume = {30},
    number = {9},
    pages = {2384-2417},
    year = {2018},
    month = {09},
    issn = {0899-7667},
    doi = {10.1162/neco_a_01112},
    url = {https://doi.org/10.1162/neco\_a\_01112},
}

@article{lyon1988,
  author={Lyon, R.F. and Mead, C.},
  journal={IEEE Transactions on Acoustics, Speech, and Signal Processing}, 
  title={An analog electronic cochlea}, 
  year={1988},
  volume={36},
  number={7},
  pages={1119-1134},
  doi={10.1109/29.1639}}

@book{Mead89_doi,
	address = {Reading, MA},
	title = {Analog {VLSI} and neural systems},
	publisher = {Addison-Wesley},
	author = {Mead, C.A.},
	year = {1989},
    doi={10.1007/978-1-4613-1639-8}
}

@InProceedings{chiavazza_2023_CVPR,
    author    = {Chiavazza, Stefano and Meyer, Svea Marie and Sandamirskaya, Yulia},
    title     = {Low-Latency Monocular Depth Estimation Using Event Timing on Neuromorphic Hardware},
    booktitle = {Proceedings of the IEEE/CVF Conference on Computer Vision and Pattern Recognition (CVPR) Workshops},
    month     = {June},
    year      = {2023},
    pages     = {4071-4080}
}

@article{gallego2020event,
  title={Event-based vision: A survey},
  author={Gallego, Guillermo and Delbr{\"u}ck, Tobi and Orchard, Garrick and Bartolozzi, Chiara and Taba, Brian and Censi, Andrea and Leutenegger, Stefan and Davison, Andrew J and Conradt, J{\"o}rg and Daniilidis, Kostas and others},
  journal={IEEE transactions on pattern analysis and machine intelligence},
  volume={44},
  number={1},
  pages={154--180},
  year={2020},
  doi={10.1109/TPAMI.2020.3008413},
  publisher={IEEE}
}

@article{greatorex_etal2024,
      title={Event-based vision for egomotion estimation using precise event timing},
      author={Hugh Greatorex and Michele Mastella and Madison Cotteret and Ole Richter and Elisabetta Chicca},
      year={2025},
      eprint={2501.11554},
      archivePrefix={arXiv},
      url={https://arxiv.org/abs/2501.11554}, 
      journal={arXiv}
}

@article{indiveri_1999,
  author={Indiveri, G. and Whatley, A.M. and Kramer, J.},
  journal={Proceedings of the Seventh International Conference on Microelectronics for Neural, Fuzzy and Bio-Inspired Systems}, 
  title={A reconfigurable neuromorphic {VLSI} multi-chip system applied to visual motion computation}, 
  year={1999},
  volume={},
  number={},
  pages={37-44},
  doi={10.1109/MN.1999.758844}}

@article{Yao_etal24,
	title = {Spike-based dynamic computing with asynchronous sensing-computing neuromorphic chip},
	volume = {15},
	issn = {2041-1723},
	url = {https://www.nature.com/articles/s41467-024-47811-6},
	doi = {10.1038/s41467-024-47811-6},
	language = {en},
	number = {1},
	urldate = {2024-07-04},
	journal = {Nature Communications},
	author = {Yao, Man and Richter, Ole and Zhao, Guangshe and Qiao, Ning and Xing, Yannan and Wang, Dingheng and Hu, Tianxiang and Fang, Wei and Demirci, Tugba and De Marchi, Michele and Deng, Lei and Yan, Tianyi and Nielsen, Carsten and Sheik, Sadique and Wu, Chenxi and Tian, Yonghong and Xu, Bo and Li, Guoqi},
	month = may,
	year = {2024},
	pages = {4464},
}

@article{Davies_etal18,
	title = {Loihi: {A} neuromorphic manycore processor with on-chip learning},
	volume = {38},
	doi = {10.1109/MM.2018.112130359},
	number = {1},
	journal = {IEEE Micro},
	author = {Davies, M. and Srinivasa, N. and Lin, T.H. and Chinya, G. and Cao, Y. and Choday, S. H. and Dimou, G. and Joshi, P. and Imam, N. and Jain, S. and Liao, Y. and Lin, C.K. and Lines, A. and Liu, R. and Mathaikutty, D. and McCoy, S. and Paul, A. and Tse, J. and Venkataramanan, G. and Weng, Y.H. and Wild, A. and Yang, Y. and Wang, H.},
	year = {2018},
	pages = {82--99},
}

@article{Bartolozzi_Indiveri07b,
	title = {Synaptic dynamics in analog {VLSI}},
	volume = {19},
	doi = {10.1162/neco.2007.19.10.2581},
	number = {10},
	journal = {Neural Computation},
	author = {Bartolozzi, C. and Indiveri, G.},
	month = oct,
	year = {2007},
	pages = {2581--2603},
}

\end{document}